\documentclass[twocolumn,showpacs,showkeys,preprintnumbers,amsmath,amssymb]{revtex4}
\usepackage{graphicx}
\usepackage{dcolumn}
\usepackage{bm}
\usepackage{epsfig}
\input{epsf}

\begin{document}


\title{Study on the large scale dynamo transition}

\author{Giuseppina Nigro}%
\author{Pierluigi Veltri}
\affiliation{%
Universit\`a della Calabria, Dipartimento di Fisica and Centro Nazionale Interuniversitario Struttura della Materia,
Unit\`a di Cosenza, I-87030 Arcavacata di Rende, Italy}%

\date{\today}

\begin{abstract}
Using the magnetohydrodynamic (MHD) description, we develop a nonlinear dynamo model 
that couples the evolution of the large scale magnetic field with turbulent dynamics of the plasma at small scale 
by electromotive force (e.m.f.). 
The nonlinear behavior of the plasma at small scale is described by using a MHD shell model 
for velocity field and magnetic field fluctuations.The shell model allow to study this problem in a large 
parameter regime 
which characterizes the dynamo phenomenon in many natural systems and which is beyond the power of 
supercomputers at today. 
Under specific conditions of the plasma turbulent state, the field fluctuations at small scales are able to trigger 
the dynamo instability. We study this transition considering the stability curve which shows a strong decrease 
in the critical magnetic Reynolds number for increasing inverse magnetic Prandlt number $\textrm{Pm}^{-1}$ 
in the range $[10^{-6},1]$ and slows an increase in the range $[1,10^{8}]$. 
We also obtain hysteretic behavior across the dynamo boundary reveling the subcritical nature of this transition. 
The system, undergoing this transition, can reach different dynamo regimes, 
depending on Reynolds numbers of the plasma flow. This shows 
the critical role that the turbulence plays in the dynamo phenomenon. 
%
%
%
In particular the model is able to reproduce the dynamical situation in which the large-scale magnetic field 
jumps between two states which represent the opposite polarities of the magnetic field, reproducing the magnetic 
reversals as observed in geomagnetic dynamo and in the VKS experiments.
%
%
\end{abstract}


\maketitle

One of the most fascinating and challenging topics in physics and astrophysics is the understanding of 
the generation and self-sustaining of magnetic fields in planets, stars, galaxies, etc. The most accredited 
mechanism is the so-called dynamo effect, i.e. the maintaining of a magnetic field against diffusive effects 
by the motion of electrically conducting fluids. This effect has also been studied in laboratory 
liquid metal experiments like Riga and VKS experiment \cite{experiment} 
%
and plays a fundamental role in our understanding of many magnetized phenomena which are of 
interest in many research fields since Faraday's study. 
%

The dynamo effect occurs 
because magnetic field lines are generally stretched at small scales by the random motion of the fluid in which they are almost `frozen'. The magnetic lines stretching leads to an exponential amplification of the field, until the backreaction (via the Lorentz force) causes the saturation of this growth. The small--scale dynamo is generated when the system reaches energy equipartition at small scales. 
During the growth of the magnetic energy at small scales, the velocity and magnetic field fluctuations interacts each other, generating an e.m.f. able to produce a large scale magnetic field. The latter increases 
due to 
the turbulent fluctuations at small scales until a saturation level is reached. 
The dynamo quenching occurs because the system attempts to conserve the total magnetic helicity when the magnetic field is growing \cite{brandenburg_PPCF09}.

In the present Letter we investigate the dynamo effect
driven solely by turbulent fluctuations, in absence of a mean flow ($\alpha^2$--dynamo), focusing 
our study on the dynamical transition towards the regime where the large scale magnetic field 
starts to be generated.  
%
Dynamo transition results from an instability: 
when, for an assigned value of the Reynolds number $Re\simeq \delta u /k_0 \nu$, the magnetic Reynolds number $\textrm{Rm} \simeq \delta u /k_0 \mu$
($\delta u$ being the r.m.s. of the turbulent velocity, $\nu$ and $\mu$ respectively the kinetic viscosity and the magnetic diffusivity) reaches a critical value $\textrm{Rm}_c$, 
the magnetic field looses its stability from a quasi zero--magnetic field state generating 
the initial exponential increasing in its values until the dynamo quenches. 
%
Generally, the dynamo bifurcation could display at least two different natures: subcritical and supercritical
 \cite{sub-super-critical}, in analogy with turbulent transitions. 
We want to investigate its nature, using a new model, recently built up, with the aim to overcome the limitation in the Reynolds number range covered by both direct numerical simulation (DNS) and by a lot of other theoretical models. 
In fact, 
although DNS are playing an important role in our understanding of the dynamo problem \citep{DNS}, 
a realistic parameter regime is still beyond the power of supercomputers at today. 
Difficulties arise from the very large values of the dimensionless parameters,
which characterize this problem in natural systems. 
\begin{figure*}[t]
\begin{center} 
\includegraphics [width=8.cm]{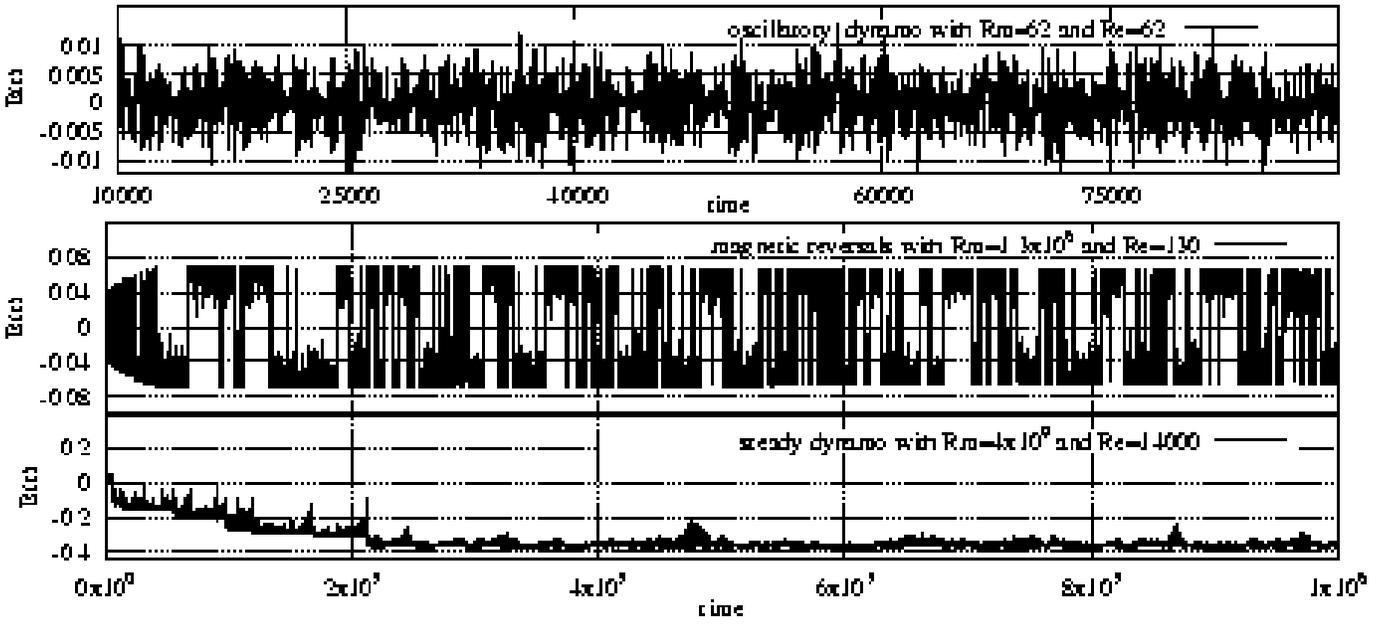}
\caption{Time evolution of the large scale magnetic field in dimensionless units in different simulations. 
The system starting from oscillatory behavior, undergoes the transition to reversals regime and finally reaches a steady dynamo for increasing values of the Reynolds numbers.
} 
\label{fig:run}
\end{center} 
\end{figure*}

In 1955, Parker \cite{parker1955} suggested that the net effect of averaging many small scale turbulent motions 
would be to produce the large scale electric field ($\alpha$ effect) generating large scale poloidal and toroidal 
magnetic field. 
The latter 
can also be generated by the differential rotation ($\alpha$-$\Omega$ effect).
In the same spirit (see also \cite{per10}), we make a decomposition of the fields in an average part, varying only on the large scale $L$, and a turbulent fluctuating part, varying at small-scales $\sim \ell$, with the assumption $\ell \ll L $ \citep{biskamp}. 
Performing this scale separation we obtain, in the induction equation at large scale, a term which describes the action of small scales on the large one consisting in a turbulent e.m.f. 
that can be written in terms of the Fourier modes of velocity (${\bf{u}}({\bf{k}},t)$) and magnetic field (${\bf{b}}({\bf{k}},t)$) small scale fluctuations as follow: 
\vglue -0.65cm
\begin{eqnarray}
\label{eq:fem}
{\bf{\epsilon}}=- \sum_{\bf {k}}  {\bf{u}} ({\bf {k}}, t) \times  {\bf{b}}^{*} ({\bf {k}}, t) \ ;
\end{eqnarray}
%
\vglue -0.25cm
this is a correlation between velocity and magnetic field fluctuations at small scales. 
%
%
Introducing a basis in the 
spectral space:
$\ \  \widehat{e_{1}} ({\bf{k}})  \ ,\ \  \widehat{e_{2}} ({\bf{k}}) =\widehat{e_{3}} ({\bf{k}}) \times \widehat{e_{1}} ({\bf{k}}) \ , \ \   \widehat{e_{3}} ({\bf{k}}) = i {\bf{k}}/\vert {\bf{k}} \vert $;
and writing expression (\ref{eq:fem}) in a form symmetric with 
respect to the change of  ${\bf{k}}$ in $- {\bf{k}}$ we finally find:
\begin{eqnarray}
\label{eq:e.m.f.}
{\bf{\epsilon}}= - \sum_{\bf k (k_z>0)}   \widehat{e_{3}} \  [ (u_1^* \ b_2 - u_2  \ b_1^*) +
(u_2^* \ b_1 - u_1 \ b_2^*)]
\end{eqnarray}
\vglue -0.25cm
where $u_1$ and $u_2$ ($b_1$ and $b_2$) are the components of ${\bf{u}}({\bf{k}},t)$ 
(${\bf{b}}({\bf{k}},t)$), along $\widehat{e_{1}}$ and $\widehat{e_{2}}$.

We describe the dynamics of the system at large scale by integrating the induction equation 
 in local approximation: we approximate the toroidal ($\widehat{e_{\varphi}}$) and the poloidal ($\widehat{e_{p}}$) unit vectors with the cartesian unit vectors, respectively $\widehat{e_{x}} $ and $\widehat{e_{z}} $. Hence the field at large scale are:
 \vglue -0.6cm
 \begin{eqnarray}
{\bf{u}}_{0} = V(y,z) \ \widehat{e_{x}} 
\;\;,\;\;\;\;
{\bf{b}}_{0}=B_{\phi}(y, t)\widehat{e_{x}} + B_{p}(y,t) \widehat{e_{z}} 
\end{eqnarray}
where ${B}_{\phi}$ and ${B}_{p}$ are the toroidal and poloidal component of the magnetic field, respectively.
We describe the dynamics at small scales by a shell model \cite{FSGC1}, 
where the wave vectors space, in which one considers 
the MHD equations, is divided into a finite number N of shells of radius $k_n= 2^n k_0$ 
(with n= 0,1,...,N; and $k_0\sim 2 \pi/\ell$ is the fundamental wave vector).
In each shell is assigned complex scalar variables $u_n(t)$ and $b_n(t)$, 
describing the dynamics of velocity and magnetic Fourier modes in the shell 
of wave vectors between $k_n$ and $k_{n+1}$. 
The nonlinear coupling of 
neighbor 
shells is chosen by preserving total energy, cross helicity, and magnetic helicity, and,  
at variance with the original MHD shell model, avoiding unphysical correlations 
of phases. 
We can write the set of self-consistent equations for our dynamo model 
in which the e.m.f. is in a form consistent with the shell model: 
%
\begin{subequations}
\begin{eqnarray}
&& 
\frac{d{B}_{\phi}}{dt} =
\frac{  i}{L} \sum_{n} ({u}_{n}^{*}  {b}_{n} - {u}_{n} {b}_{n}^{*})+
{B_{p} \frac{V}{L} } - \mu \frac{{B}_{\phi}}{L^{2}}  \ ,
\label{eq:Bphi2}
\\
&& 
\frac{d{B}_{p}}{dt} =
\frac{  i}{L} \sum_{n}  ({u}_{n}^{*} {b}_{n} - {u}_{n} {b}_{n}^{*} ) 
 - \mu \frac{{B}_{p}}{L^{2}} \ ,
\label{eq:Bp2}
\\
&& 
{\frac{d{u_{n}}}{dt}} = 
 k_{n} ({B}_{\phi} + {B}_{p}) b_{n}+ {\rm i}k_n \Big[(u_{n+1}^*u_{n+2}-b_{n+1}^*b_{n+2})+
\nonumber
\\
&&
-\frac{1}{4}(u_{n-1}^*u_{n+1}-b_{n-1}^*b_{n+1})+\frac{1}{8}(u_{n-2}u_{n-1}-b_{n-2}b_{n-1})\Big]
\nonumber
\\
&&
 \;\;\;\;\;\;\;\;\;\;\;\;\;\;\; - \nu k_n^2 u_n + f_n,
\label{eq:sabrav}
\\
&&
{\frac{d{b_{n}}}{dt}}= 
i k_{n} ({B}_{\phi} + {B}_{p}) u_{n} +  \frac{i k_n}{6} \Big[(u_{n+1}^*b_{n+2}-b_{n+1}^*u_{n+2})
\nonumber
\\
&&
+(u_{n-1}^*b_{n+1}-b_{n-1}^*u_{n+1})-(u_{n-2}b_{n-1}-b_{n-2}u_{n-1})\Big]  
\nonumber
\\
&&
 \;\;\;\;\;\;\;\;\;\;\;\;\;\;\; - \mu k_n^2 b_n;
\label{eq:sabrab}
\end{eqnarray}
\end{subequations}
%
where the spatial derivative associated at large scale is estimated dividing by the typical large scale L; 
$\nu$ is the viscosity and $\mu$ is the diffusivity of the MHD flow; 
$n$ is the shell number ($n=0,...N$); 
finally $f_n$ is an external forcing term applied on the first shell ($n=0$). 
This forcing, that drives the fluid flow to become unstable, has the property to preserve the energy flux 
to small scales. 
This is the same forcing used in \cite{nigro10} with correlation time equal to 1. 
%
The first terms in the RHS of Eqs. (\ref{eq:sabrav})-(\ref{eq:sabrab}) 
describe the effect of the large scale magnetic field on the small--scale turbulent dynamics. 
Due to these terms, the turbulence becomes anisotropic because the field fluctuations self organize 
themselves in coherent structures, 
channeled along the large scale magnetic field $\bf b_0$ as Alfv\'en waves.
\begin{figure}[t]
\begin{center} 
\includegraphics [width=8.3cm]{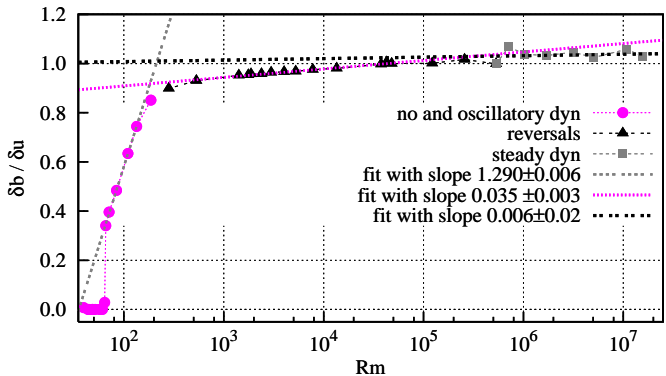} 
\caption{(Color online) Dynamical behavior of $\delta b / \delta u$ for increasing $\textrm{Rm}$. 
Starting from no magnetic field state, we check the values of $\delta b / \delta u$ during the 
running time of the simulation in which we increase $\textrm{Rm}$ for $\nu=10^{-5}$. 
During this time the system undergoes dynamo transition 
(magenta circles) reaching oscillatory regime, afterwards the system goes towards reversals regime (black triangles) coming to steady states where the large scale magnetic field saturates (gray squares).
}
\label{fig:regimes} 
\end{center} 
\end{figure}

We have obtain the e.m.f. consistently with shell model without linear approximation.
The e.m.f. tends to vanish when the system evolves towards a state 
of strong correlation between velocity and magnetic field, which characterizes the Alfv\'enic subspaces. 
This is an attractor of the dynamics of the system \cite{dobrowolny80}.
We solve the model Eqs. assuming $V=0$, that is equivalent to solve the dynamo problem for 
Rossby number $\textrm{Ro} = \frac{\delta u}{V} \frac{\delta b}{B_p} \gg 1$. 
Hence ${B}_p(t)={B}_{\phi}(t)=\textrm{B}(t)$ and the model describes $\alpha^2$ dynamo problem, 
in which the shear due to the differential rotation at large scale is neglected. 
Even in the absence of the macroscopic shear, the $\alpha$ effect can give rise 
to dynamo action. This effect seems to play a decisive role for 
planetary magnetic fields as geomagnetic field. It can be very strong in the fully convective
stars as the late-type stars and it may provide a possible mechanism to explain magnetic activity, along 
with nonaxisymmetric field as observed in many active stars \cite{meinel}.

The model Eqs. are numerically solved by using a fourth order Runge-Kutta scheme.
The results are in dimensionless units: the field fluctuations are measured in Alfv\'en velocity unit $c_A$, 
the time in eddy-turn-over time ($1/(k_0 u_0)$), the lengths are normalized to $1/k_0$, and finally the 
dissipative coefficients are normalized to $c_A / k_0$. 

At the beginning of each simulation, we let the system become turbulent at small scales, i.e. we keep 
$\textrm{B}=0$ up to $1000$ unit times. During this time the kinetic and the magnetic energy at 
small scales grow forming a power law spectrum.
The system tends to the energy equipartition at small scales, exactly achieved when $\nu=\mu$. 
After that, when we are sure that the turbulence is fully developed, we introduce a magnetic field seed 
of amplitude $10^{-10}$ at large scale and we check whether the dynamo effect starts to develop.

\begin{figure}[t]
\begin{center} 
\includegraphics [width=8.3cm]{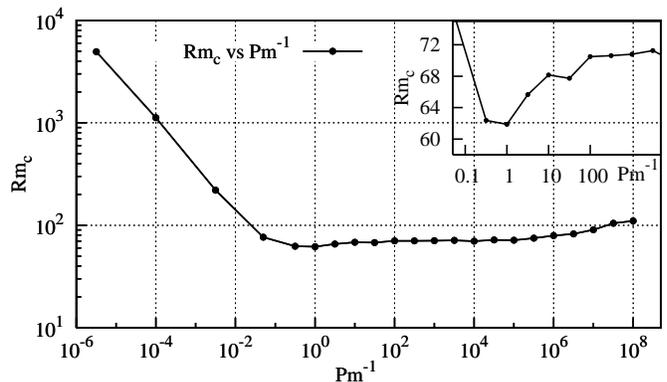} 
\caption{
The stability curve $\textrm{Rm}_c$ vs $\textrm{Pm}^{-1}$ in log scale. 
The inset in semi--log scale shows the slight increase of $\textrm{Rm}_c$ for increasing $\textrm{Pm}^{-1}>1$.
}
\label{fig:stability_curve} 
\end{center} 
\end{figure}
The numerical results reveal a strong sensitivity of the system with respect to the magnetic 
Reynolds number $\textrm{Rm}$ and a dependence on the hydrodynamic Reynolds number 
$\textrm{Re}$. Depending on these parameters, the system evolves towards different scenarios: 
i) no dynamo; ii) oscillatory dynamo; iii) magnetic reversals; iv) steady dynamo. 
The transition from oscillatory dynamo to a steady dynamo, going through a reversals regime, 
is obtained by increasing $\textrm{Re}$ and $\textrm{Rm}$ (Fig.~\ref{fig:run}), reproducing in this way 
the same dynamo regimes as observed in the liquid metal experiments \cite{experiment}. 
%

In order to better characterize the different regimes, we have performed a simulation with a fixed value of the kinematic viscosity but changing the value of the magnetic diffusivity after a time interval sufficiently long with respect to the dynamical evolution times and we have studied how the two parameters $\delta b/\delta u$ and 
$rms(\textrm{B})/\delta u$ change with the magnetic Reynolds number $\textrm{Rm}$.

The $\delta u / \delta b $ function illustrated in Fig.~\ref{fig:regimes} for increasing \textrm{Rm} 
shows a zero order discontinuity corresponding to dynamo transition (the step for $\textrm{Rm}=67$) and a first order discontinuity corresponding to the transition from oscillatory dynamo to reversals regime. The kind of these 
discontinuities marks the different nature of the transitions, in particular we can argue that the dynamo 
effect occurs as a plasma instability while the others transitions occur with continuity.

Our model allows us to study the dynamo instability inside a very large values of critical 
parameters not accessible by DNS yet, but more realistic for the astrophysical objects. 
We have investigated the dynamo threshold for different Prandlt number, 
reproducing the stability curve in a very wide range of values, illustrated in Fig.~\ref{fig:stability_curve}, 
which shows a little dependence of $\textrm{Rm}_c$ for $\textrm{Pm}^{-1}>1$; 
and a strong monotonic increase of $\textrm{Rm}_c$ for decreasing $\textrm{Pm}^{-1}<1$.
The reason of this strong increase could be related to the need to have small 
magnetic diffusivity to compensate the large viscosity 
in order to have small scales field fluctuations large enough, to consistently trigger the e.m.f. 
at large scale for the dynamo onset.
On one hand, this result is consistent with what was obtained by other models in a smaller range 
of critical parameters: $\textrm{Pm}^{-1} \in [1,100])$ 
\cite{stability-curve}, giving us the possibility to be confident on our model, 
and on the other hand, it extends these results in larger range of values not investigated yet.  
%
The model hence
reaches values more realistic for the 
astrophysical objects. For instance, for interstellar or intracluster medium $\textrm{Pm}\gg1$, 
in the Sun's convective zone, $\textrm{Pm} \sim 10^{-7}$ to $\sim 10^{-4}$, in planets 
$\textrm{Pm}\sim 10^{-5}$ and in protostellar disks $\textrm{Pm} \ll 1$.
%
%
%

The dynamo onset could take place in subcritical or supercritical way. 
In some models and experiments it is subcritical, as it has showed in 
some small--scale dynamos, revealing a hysteretic behavior \cite{hysteresis}.
\begin{figure}[t]
\begin{center} 
\includegraphics [width=8.3cm]{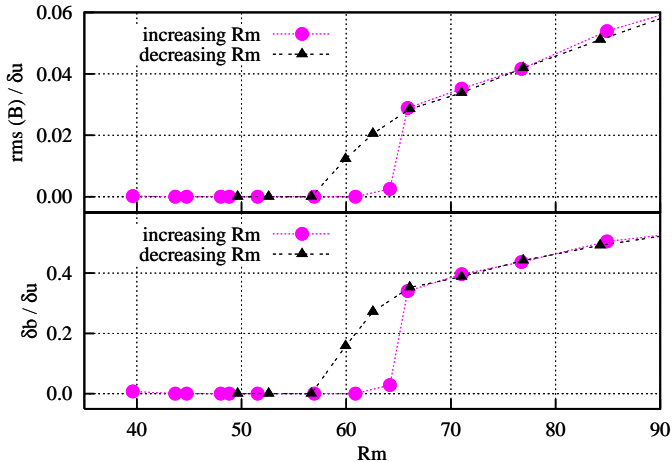}
\caption{(Color online) Hysteresis cycles of $\textrm{rms}(\textrm{B})/ \delta u$ (top panel) 
and $\delta b / \delta u$ (bottom panel) obtained around the stability margin of the bifurcation for the simulation  
with $\nu=10^{-5}$ and changing $\textrm{Rm}$: 
we increase $\textrm{Rm}$ 
reaching dynamo onset in which the magnetic field displays oscillations (first phase: magenta circles), 
afterwards we decrease $\textrm{Rm}$ (second phase: black triangles).
}
\label{fig:hysteresis} 
\end{center} 
\end{figure}
In order to investigate the nature of this instability as described in our model, we have reproduced hysteresis 
cycles realized as following: for a fixed value of $\nu$, starting from a state of quasi zero--magnetic field 
characterized by low $\textrm{Rm}$, we have dynamically increased $\textrm{Rm}$ during the 
simulation time in order to destabilize the system from the initial state. 
During the simulation we have checked the time evolution of the order parameters: 
the ratio between the r.m.s. of $\textrm{B}$ over $\delta u$ and $\delta b$ over $\delta u$.
Only after a critical value of $\textrm{Rm}$, corresponding to the threshold for 
the dynamo action to set up, we have found the increase of the large scale
magnetic field (first phase). Once the dynamo sets in, we have decreased $\textrm{Rm}$ (second phase). 
In the second phase, order parameters not follow the path inverse to the first phase, realizing 
a hysteresis loop. We have investigated hysteresis cycles considering a wide range of values of $\nu$ 
down to $10^{-11}$. 
For the example reported in Fig.~\ref{fig:hysteresis} where $\nu=10^{-5}$, 
starting from a quasi null magnetic field, we have increased $\textrm{Rm}$ up to $\textrm{Rm}_c=67$, 
corresponding to the dynamo threshold, by finding for the order parameter the values 
shown by the magenta circles in Fig.~\ref{fig:hysteresis}. 
After the dynamo onset, keep increasing $\textrm{Rm}$, the system displays magnetic oscillation and subsequently magnetic reversals. Once reached these regimes, we have decreased $\textrm{Rm}$. 
The values of the order parameters, representing now by the black triangles, 
lie on the same curve when the system is in latter regimes, afterwards the order parameters lie on a
different curve starting from $\textrm{Rm}_c$ to lower values. 
This shows the hysteretical behavior of the transition and its subcritical nature. 
We have found also a slight suppression of hydrodynamic turbulence during the transition \cite{cattaneo96}, 
revealed by the decrease of hydrodynamic Reynolds number $\textrm{Re} \simeq \delta u/ k_0 \nu$. 
In the example in Fig.~\ref{fig:hysteresis}, $\textrm{Re}$, starting from a value around $8000$, 
slightly decreases to $7000$ when the dynamo action is developed. 
We can argue that the subcritical nature of dynamo bifurcation could come out from the reduction of turbulence 
due to the magnetic field growing, which ensure the maintain 
of the dynamo effect for lower $\textrm{Rm}$ values. 
%

We present a dynamo model developed for the large scale magnetic field. 
The model solves the induction equation in local approximation at large scale, 
while the turbulent dynamics at small scales is described by a MHD shell model.  
We point out that there is no prescribed form (like $\alpha$-term) for e.m.f., 
but it is consistently described by the analytical form of the shell model. 
Moreover at variance with others dynamo shell models \cite{nigro10,benzi10} which use \textit{ad hoc} terms 
to mimic the dynamo instability, this model is able to reproduce different regimes in dynamical way 
without any \textit{ad hoc} assumption.
An important result of this research is related to the critical role that the turbulence plays in the dynamo 
phenomenon showing how different regimes are obtained depending on Reynolds numbers of the plasma flow. 
This property of the model is also confirmed by laboratory experiments inside the Reynolds number accessible 
to these. 
One of the strength of our model, due to the use of the shell technique, consists of the capability to describe 
the dynamo transition and the different regimes where it can act in a very large values of the Reynolds numbers. 
The study on the dynamo transition shows its subcricital nature supported by hysteresis cycles around the 
stability margin of the bifurcation.  
The model reproduces results consistent with other dynamo models in the parameter range accessible to these, 
therefore showing its reliability. Moreover it makes a step forward in our understanding of the dynamo 
problem due to its capability to cover a larger range of values not yet accessible by others models and  
by DNS, but more realistic for a natural dynamo processes. Finally we consider the results here obtained a 
good stimulus for both DNS and experimental studies on this problem.
%
%
%

%
\end{document}